\documentclass[11pt]{article}
\usepackage{graphicx,verbatim,array,multicol,palatino,amssymb}
\def\thick#1{\hbox{\rlap{$#1$}\kern0.25pt\rlap{$#1$}\kern0.25pt$#1$}}
\def\jump{\vskip3mm\noindent}
\def\bib{\vskip12pt\par\noindent\hangindent=1 true cm\hangafter=1}
\def\real{{\mathbb R}}
\def\etal{{\em et al.}{}}
\def\bB{{\bf B}}
\def\bOmega{{\thick\Omega}}
\def\fhatO{\fhat_{\scriptscriptstyle O}}
\def\bnuhatO{\bnuhat_{\scriptscriptstyle O}}
\def\transpose{T}
\def\trans{^{\transpose}}
\def\by{{\bf y}}
\def\fhat{{\widehat f}}
\def\bnuhat{{\widehat\bnu}}
\def\bnu{{\thick\nu}}
\def\bBtilde{{\widetilde \bB}}
\def\bw{{\bf w}}
\def\xtilde{{\widetilde x}}
\def\bkappa{{\thick\kappa}}
\def\bxtilde{{\widetilde \bx}}
\def\bx{{\bf x}}
\def\bD{{\bf D}}
\def\fhatP{\fhat_{\scriptscriptstyle P}}
\def\bnuhatP{\bnuhat_{\scriptscriptstyle P}}
\def\fhatS{\fhat_{\scriptscriptstyle S}}
\def\bbetahat{{\widehat\bbeta}}
\def\buhat{{\widehat \bu}}
\def\bbeta{{\thick\beta}}
\def\bu{{\bf u}}
\def\bX{{\bf X}}
\def\bZ{{\bf Z}}
\def\bvarepsilon{{\thick\varepsilon}}
\def\bzero{{\bf 0}}

\def\sigu{\sigma_u}
\def\bI{{\bf I}}
\def\bnuhatO{\bnuhat_{\scriptscriptstyle O}}
\def\bC{{\bf C}}
\def\bL{{\bf L}}
\def\bOmega{{\thick\Omega}}
\def\bU{{\bf U}}
\def\diag{\mbox{diag}}
\def\bd{{\bf d}}
\def\bdZ{\bd_{\scriptscriptstyle Z}}
\def\bUZ{\bU_{\scriptscriptstyle Z}}
\def\bUX{\bU_{\scriptscriptstyle X}}
\def\SBMD{{\tt SBMD}}
\def\sigeps{\sigma_{\varepsilon}}
\def\Cov{\mbox{Cov}}
\def\logit{\mbox{logit}}
\def\bmath#1{\mbox{\boldmath$#1$}}
\def\beps{\bmath{\varepsilon}}
\def\bZspline{\bZ_{\mbox{{\tiny spline}}}}
\def\bZsubj{\bZ_{\mbox{{\tiny subj}}}}
\def\Rhat{{\widehat R}}
\def\myand{and\ }
\def\BachrachHastieWangNarasimhanMarcus#1 % 1999
{
\bib
Bachrach, L.K., Hastie, T., Wang, M.-C., Narasimhan, B.
\myand Marcus, R. (#1).
Bone mineral acquisition in healthy Asian, Hispanic, Black and
Caucasian youth. A longitudinal study. 
{\it Journal of Clinical Endocrinology and Metabolism},
{\bf 84}, 4702--12.
}

\def\Berndt#1 % 1991
{\bib
Berndt, E. R. (#1).
{\it The Practice of Econometrics: {C}lassical and Contemporary.}
Reading, Massachusetts: Addison-Wesley.
}

\def\EilersMarx#1
{\bib
Eilers, P.H.C. \myand Marx, B.D. (#1).
Flexible smoothing with B-splines and penalties
(with discussion).
{\it Statistical Science}, {\bf 11}, 89--121.
}

\def\CarrollRuppertStefanskiCrainiceanu#1 % 2006
{\bib
Carroll, R.J., Ruppert, D., Stefanski, L.A. 
and Crainiceanu, C.M. (#1).
{\it Measurement Error in Nonlinear Models
(Second Edition)}. Boca Raton, Florida: Chapman \& Hall/CRC.
}

\def\CantoniHastie#1 % 2002
{\bib
Cantoni, E. and Hastie, T. (#1).
Degrees of freedom tests for smoothing splines.
{\it Biometrika}, {\bf 89}, 251--265.
}

\def\ChaudhuriMarron#1 % 1999
{\bib
Chaudhuri, P. \& Marron, J.S. (#1).
SiZer for exploration of structures in curves.
{\it Journal of the American Statistical Association},
{\bf 94}, 807--823.
}

\def\CrainiceanuRuppertWand#1 % 2005
{\bib
Crainiceanu, C., Ruppert, D. and Wand, M.P. (#1).
Bayesian analysis for penalized spline regression
using WinBUGS. {\it Journal of Statistical Software}, 
Volume 14, Issue 14.
}

\def\deBoor#1 % 1978
{\bib
de Boor, C. (#1). 
{\it A Practical Guide to Splines.}
Berlin: Springer-Verlag.
}

\def\DenisonHolmesMallickSmith#1 % 2002
{\bib
Denison, D.G.T., Holmes, C.C., Mallick, B.K. and Smith, A.F.M. (#1).
{\it Bayesian Methods for Nonlinear Classification and Regression}.
Chichester, UK: Wiley.
}

\def\EubankTAS#1  % 1994
{\bib
Eubank, R.L. (#1).
A simple smoothing spline. 
{\it The American Statistician}, {\bf 48},
103--106. 
}

\def\EubankBOOK#1 % 1999
{\bib
Eubank, R. L. (#1).
{\it Nonparametric Regression and Spline Smoothing.}
New York: Marcel Dekker.
}

\def\GreenSilverman#1 % 1994
{\bib
Green, P.J. \myand Silverman, B.W. (#1). {\it Nonparametric Regression
and Generalized Linear Models.} London: Chapman and Hall.
}

\def\GuANOVA#1 % 2002
{\bib
Gu, C. (#1).
{\it Smoothing Spline ANOVA Models}.
New York: Springer.
}

\def\HallOpsomer#1 % 2005
{
\bib
Hall, P. and Opsomer, J.D. (#1). Theory for
penalised spline regression.  {\it Biometrika},
{\bf 92}, 105--118.
}

\def\HannigMarron#1 % 2006
{
\bib
Hannig, J. \myand Marron, J.S. (#1). Advanced distribution theory
for SiZer. {\it Journal of the American Statistical Association},
{\bf 101}, 484--499.
}

\def\HastieCRAN#1 % 2006
{\bib
Hastie, T. (#1).
{\tt gam 0.97}. R package. {\tt http://cran.r-project.org}.
}

\def\HastieTibshiraniFriedman#1 % 2001
{\bib
Hastie, T., Tibshirani, R. \myand Friedman, J. (#1).
{\it The Elements of Statistical Learning.}
New York: Springer-Verlag.
}

\def\LuoWahba#1 % 1997
{\bib
Luo, Z. and Wahba, G. (#1).
Hybrid adaptive splines. {\it Journal of the
American Statistical Association}, {\bf 92},
107--116.
}

\def\NgoWand#1 %2003
{\bib
Ngo, L. and Wand, M.P. (#1). 
Smoothing with mixed model software.
{\it Journal of Statistical Software},
Volume 9, Issue 1.
}

\def\Nussbaum#1 % 1985
{\bib
Nussbaum, M. (#1).
Spline smoothing in regression models and
asymptotic efficiency in $L_2$.
{\it The Annals of Statistics}, {\bf 13},
984--997.
}

\def\NychkaCummins#1 % 1996
{\bib
Nychka, D. and Cummins, D.J. (#1).
Comment on paper by Eilers and Marx.
{\it Statistical Science}, {\bf 11}, 104--105.
}

\def\OSullivanStatSci#1 % 1986
{\bib
O'Sullivan, F. (#1).
A statistical perspective on ill-posed inverse problems (with discussion). 
{\it Statistical Science},
{\bf 1}, 505--527.
}

\def\RuppertKNOTS#1 % 2002
{\bib
Ruppert, D. (#1).
Selecting the number of knots for penalized splines.
{\it Journal of Computational and Graphical Statistics},
{\bf 11}, 735--757.
}

\def\RuppertWandCarroll#1 % 2003
{\bib
Ruppert, D., Wand, M. P. \myand Carroll, R.J. (#1). 
{\it Semiparametric Regression}.
New York: Cambridge University Press.
}

\def\SchoenbergGRAD#1 % 1964
{\bib
Schoenberg, I.J. (#1). 
Spline functions and the problem of gradation.
{\it Proceedings of the National Academy of Sciences
of the United States of America}, {\bf 52}, 947--950.
}

\def\Solo#1 % 2000
{\bib
Solo, V. (#1).
A simple derivation of the smoothing spline.
{\it The American Statistician}, {\bf 54}, 40--43.
}

\def\Speed#1
{\bib
Speed, T. (#1). Comment on paper by Robinson.
{\it Statistical Science}, {\bf 6}, 42--44.
}

\def\SpiegelhalterThomasBest#1 % 2000
{\bib
Spiegelhalter, D., Thomas, A. and Best, N. (#1).
WinBUGS Version 1.3 User Manual. 
{\tt www.hrc-bsu.cam.ac.uk/bugs}.
}

\def\Verbyla#1 %1994 
{\bib
Verbyla, A.P. (#1).
Testing linearity in generalized linear models.
{\it Contributed Paper, 17th International Biometrics Conference, 
Hamilton, Canada.}, 177.
}

\def\WahbaSIAM#1
{\bib
Wahba, G. (#1). {\it Spline Models for Observational Data.}
Philadelphia: SIAM.
}

\def\WandREGSPL#1 % 2000
{\bib
Wand, M. P. (#1).
A comparison of regression spline smoothing procedures.
{\it Computational Statistics}, {\bf 15}, 443--462.
}

\def\WelhamCullisKenwardThompson#1 % 2007
{\bib
Welham, S.J., Cullis, B.R., Kenward, M.G. and Thompson, R. (#1).
A comparison of mixed model splines for curve fitting.
{\it Australian and New Zealand Journal of Statistics},
{\bf 49}, 1--23.
}

\def\WhittakerRobinson#1 % 1967
{\bib
Whittaker, E.T. \myand Robinson, G. (#1). The Newton-Cotes formulae
of integration. Section 76 in 
{\it The Calculus of Observations: A Treatise on Numerical Mathematics},
4th Edition. New York: Dover, pp.152--156.
}

\def\ZhaoStaudenmayerCoullWand#1 % 2006
{\bib
Zhao, Y., Staudenmayer, J., Coull, B.A. and Wand, M.P. (#1).
General design Bayesian generalized linear mixed models.
{\it Statistical Science}, {\bf 21}, 35--51.
}

\begin{document}
%
%\null\vskip2cm

\begin{center}
{\LARGE\bf On semiparametric regression with O'Sullivan penalised splines}

\vskip3mm
{\sc By M. P. Wand}
\vskip3mm
{\it School of Mathematics and Applied Statistics,
University of Wollongong, Wollongong 2522, Australia}
\vskip3mm
{\sc and J.T. Ormerod}
\vskip3mm
{\it School of Mathematics and Statistics,
University of New South Wales, Sydney 2052, Australia}

\vskip1mm

\centerline \date{25th June, 2007}
\vskip3mm
{\sc Summary}\\[1.5ex]
\end{center}

\noindent
This is an expos\'e  on the use of O'Sullivan penalised splines
in contemporary semiparametric regression, 
including mixed model and Bayesian formulations. 
O'Sullivan penalised splines are similar to P-splines,
but have an advantage of being a direct generalisation of smoothing
splines. Exact expressions for 
the O'Sullivan penalty matrix are obtained.
Comparisons between the two reveals that O'Sullivan
penalised splines more closely mimic the natural boundary
behaviour of smoothing splines.  
Implementation in modern computing environments such as
{\tt Matlab}, {\tt R} and {\tt BUGS} is discussed.

\jump

\vskip5mm

\section{Introduction}\label{sec:intro}

Splines continue to play a central role in 
nonparametric and semiparametric regression 
modelling. Recent synopses include Eubank (1999),
Gu (2002),  Ruppert, Wand \& Carroll (2003)
and Denison, Holmes, Mallick \& Smith (2002).
In all but the last reference, smooth functional
relationships are fitted using a large basis of
spline functions subject to penalisation. Up
until the mid-1990s most literature on spline-based 
nonparametric regression was concerned with {\em smoothing splines},
and their multivariate extension {\em thin plate splines},
where the penalty takes a particular form and
the number of basis functions roughly equals the sample
size (e.g. Wahba, 1990; Green \& Silverman, 1994).
However, in recent years, there has been
a great deal of research on more general spline/penalty
strategies, most of which use considerably fewer basis
functions. Driving forces include: 
\begin{itemize}
\item more complicated models, often with several smooth
functions;
\item larger data sets, where smoothing and thin plate splines
become computationally intractable,
\item mixed model and Bayesian representations of smoothers
that lend themselves to the use of established software, such as 
{\tt BUGS},  {\tt lme()} in {\tt R} and {\tt PROC MIXED} in {\tt SAS}; 
provided the number of basis functions is
relatively low.
\end{itemize}
Ruppert, Wand \& Carroll (2003) summarise and provide access
to many of these developments. The term {\em penalised splines} 
has emerged as a descriptor for general spline fitting subject 
to penalties.

O'Sullivan (1986, Section 3) introduced a class of penalised splines 
based on B-spline basis functions. O'Sullivan penalised splines 
are a direct generalisation of smoothing splines 
in that the latter arises when the maximal number of B-spline
basis functions are included. Like smoothing splines, O'Sullivan
penalised splines possess the attractive feature of 
natural boundary conditions 
(e.g. Green \& Silverman, 1994, p.12). 
They have also become the most widely used
class of penalised splines in statistical analyses
due to their implementation in the popular {\tt R} and 
{\tt S-PLUS} function {\tt smooth.spline()} and associated generalised
additive model software (e.g. the {\tt gam} library in {\tt R};
Hastie, 2006).

Despite the omnipresence of O'Sullivan penalised splines,
their use in semiparametric regression contexts, particularly
those involving mixed model and Bayesian representations,
is not very common. Recently, Welham, Cullis, Kenward \& Thompson (2007)
showed how most of the commonly used penalised splines 
can be treated within a single mixed model framework, 
although they did not work explicitly with the form 
given in O'Sullivan (1986).

Our contributions in this paper are 
(a) providing an exact matrix expression 
for the penalty of O'Sullivan splines
that allows implementation in a few lines of a matrix-based
computing language, 
(b) comparison with their closest penalised spline
relative, P-splines (Eilers \& Marx, 1996),
which reveal some noticeable differences near the boundaries,
(c) demonstrate explicitly,
including with {\tt R} code, how O'Sullivan splines can be 
simply added to the mixed model-based regression
armoury, and 
(d) investigate their efficacy in Bayesian
semiparametric regression using MCMC software such as 
{\tt BUGS} and its variants.
We conclude that the several attractive features of O'Sullivan 
penalised splines --- smoothness, numerical stability, natural 
boundary properties, direct generalisation of smoothing splines 
--- makes them a very good choice of basis in semiparametric regression.

Section \ref{sec:osu} provides a brief description of O'Sullivan
penalised splines. Comparison with P-splines is made in
Section \ref{sec:comp}. Section \ref{sec:mix} describes
mixed model representation of O'Sullivan penalised splines
and how they can be used in models that benefit from this
representation. Issues concerning Bayesian penalised spline
smoothing and Markov chain Monte Carlo are described in
Section \ref{sec:baye}. O'Sullivan splines of general
degree are described in Section \ref{sec:hdOsul}.
Closing discussion is given in Section \ref{sec:discn}. 
An appendix contains relevant {\tt R} code.

\section{O'Sullivan Penalised Splines}\label{sec:osu}

O'Sullivan penalised splines have already been
described several times in the literature. 
A recent reference is the Chapter 5 Appendix of 
Hastie, Tibshirani \& Friedman (2001). 
A brief sketch is given here for convenience.

Consider the simplest nonparametric regression setting
\begin{equation}
y_i=f(x_i)+\varepsilon_i,\quad 1\le i\le n,
\label{eq:simpNR}
\end{equation}
where $(x_i,y_i)\in\real\times\real$. 
Suppose that an estimate of $f$ is required over
$[a,b]$, an interval containing the $x_i$'s.
For an integer $K\le n$ let $\kappa_1,\ldots,\kappa_{K+8}$ 
be a knot sequence such that 
$$a=\kappa_1=\kappa_2=\kappa_3=\kappa_4<\kappa_5<\cdots<\kappa_{K+4}
<\kappa_{K+5}=\kappa_{K+6}=\kappa_{K+7}=\kappa_{K+8}=b$$
and let $B_1,\ldots,B_{K+4}$ be the cubic B-spline
basis functions defined by these knots
(see e.g. pp.160--161 of Hastie \etal, 2001).
Set up the $n\times (K+4)$ design matrix $\bB$
with $(i,k)$ entry $B_{ik}=B_k(x_i)$ and the 
$(K+4)\times (K+4)$ penalty matrix $\bOmega$ with $(k,k')$
entry
$$\bOmega_{kk'}=\int_a^b B''_k(x)B_{k'}''(x)\,dx.$$
Then an estimate of $f$ at location $x\in\real$ can be obtained
as 
\begin{equation}
\fhatO(x;\lambda)\equiv\bB_x\bnuhatO\quad\mbox{where}\quad
\bnuhatO\equiv(\bB\trans\bB+\lambda\bOmega)^{-1}\bB\trans\by,
\label{eq:fhat.osu}
\end{equation}
$\bB_x\equiv[B_1(x),\ldots,B_{K+4}(x)]$ and $\lambda>0$
is a smoothing parameter.

\begin{figure}[ht]
\null\centerline{\includegraphics[width=12cm]{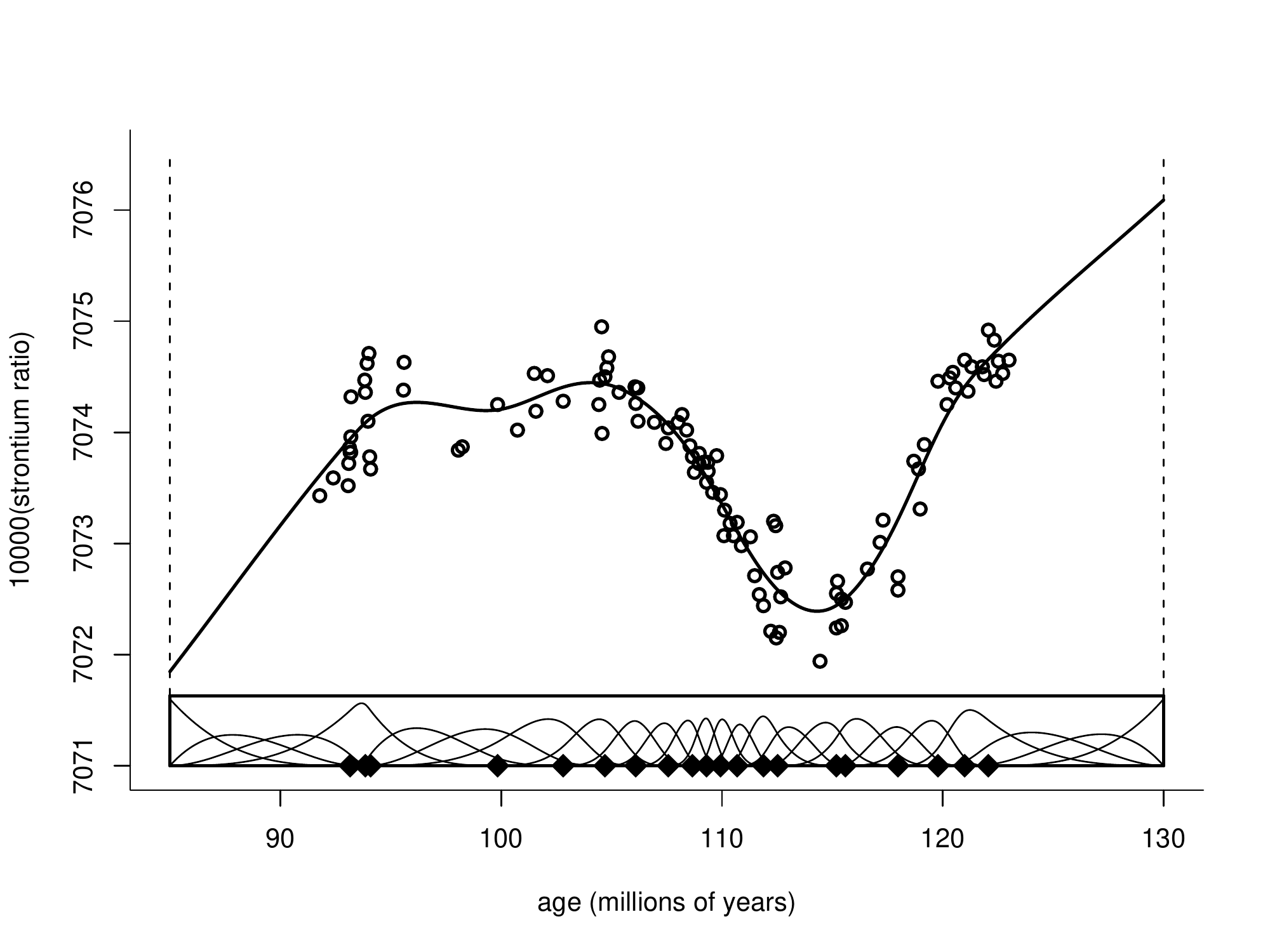}}
\caption
{\it
Illustration of natural boundary properties of a 
20-interior knot O'Sullivan penalised spline fit to the fossil data
over the interval $[85,130]$ millions of years. The interior knots
are shown as solid diamonds ($\blacklozenge$). Inset:
The 24 B-spline basis functions.
}
\label{fig:nbc}
\end{figure}

Note that the cubic smoothing spline arises
in the special case 
$K=n$ and
$\kappa_{k+4}=x_k$, $1\le k\le n$, provided the 
$x_i$'s are distinct
(e.g. Green \& Silverman, 1994, Section 3.6).
Apart from giving a smooth (twice continuously differentiable)
scatterplot smooth, $\fhatO(\cdot;\lambda)$ has good numerical
properties. The basis functions are bounded and so not prone to
overflow problems. Moreover, $\bB\trans\bB$ is 4-banded,
which leads to $O(n)$ algorithms when $K$ is close to $n$ 
(e.g. Hastie, \etal, 2001). In addition, $\fhatO(\cdot;\lambda)$
satisfies so-called natural boundary conditions, meaning that 
$$\fhatO''(a;\lambda)=\fhatO'''(a;\lambda)=
\fhatO''(b;\lambda)=\fhatO'''(b;\lambda)=0$$
and implying that $\fhatO(\cdot;\lambda)$ is linear over 
$[a,\kappa_5]$ and $[\kappa_{K+4},b]$. Figure \ref{fig:nbc}
illustrates these natural boundary properties of $\fhatO(\cdot;\lambda)$
for data on ratios of strontium isotopes found in fossil 
shells and their age; see Chaudhuri \& Marron (1999) for
details. Also, $\fhatO(\cdot;\lambda)$ approximates the
least squares line as $\lambda\to\infty$. The implication for
mixed model smoothing is that the induced fixed effects
component corresponds to straight line basis functions.
Details are given in Section \ref{sec:mix}.

Computation of the design matrix $\bB$ is usually quite
easy. For example, B-splines are readily available in the {\tt Matlab},
{\tt R} and {\tt S-PLUS}
computing environments. Otherwise recurrence formulae 
(e.g. de Boor, 1978; Eilers \& Marx, 1996) can be called upon.
However, computation of $\bOmega$ requires some additional effort.
In Section \ref{sec:hdOsul}, while treating general degree
O'Sullivan penalised splines, we derive an exact matrix algebraic
expression for the corresponding penalty matrices. In the
cubic case our theorem reduces to
the expression:
\begin{equation}
\bOmega=(\bBtilde'')^T\mbox{\rm diag}(\bw)\bBtilde''
\label{eq:OmegaExact}
\end{equation}
where $\bBtilde''$ is the $3(K+7)\times(K+4)$ 
matrix with $(i,j)$ entry
$B''_j(\xtilde_i)$, $\xtilde_i$ 
is the $i$th entry of the vector
$$\bxtilde=\left(\kappa_1,\frac{\kappa_1+\kappa_2}{2},\kappa_2,
                 \kappa_2,\frac{\kappa_2+\kappa_3}{2},\kappa_3,
\ldots,\kappa_{K+7},\frac{\kappa_{K+7}+\kappa_{K+8}}{2},\kappa_{K+8}\right).$$
and $\bw$ is the $3(K+7)\times1$ vector given by
$$\bw=\left(\textstyle{\frac{1}{6}}(\Delta\bkappa)_1,
\textstyle{\frac{4}{6}}(\Delta\bkappa)_1,
\textstyle{\frac{1}{6}}(\Delta\bkappa)_1,
\textstyle{\frac{1}{6}}(\Delta\bkappa)_2,
\textstyle{\frac{4}{6}}(\Delta\bkappa)_2,
\textstyle{\frac{1}{6}}(\Delta\bkappa)_2,\ldots,
\textstyle{\frac{1}{6}}(\Delta\bkappa)_{K+7},
\textstyle{\frac{4}{6}}(\Delta\bkappa)_{K+7},
\textstyle{\frac{1}{6}}(\Delta\bkappa)_{K+7}\right),
$$
where $(\Delta\bkappa)_k\equiv\kappa_{k+1}-\kappa_k$, $1\le k\le K+7$.
\noindent
Result (\ref{eq:OmegaExact}) is none other than Simpson's rule applied
over each of the inter-knot differences. This is because each 
$B_i''B_j''$ function is piecewise quadratic.
For commonly used values of $K$,
(\ref{eq:OmegaExact}) allows straightforward computation of $\bOmega$ 
in matrix-based  languages such as {\tt Matlab}, {\tt R} and 
{\tt S-PLUS}. In the Appendix we demonstrate computation of 
$\bOmega$ in 4 lines of {\tt R} code.

Lastly, we mention knot choice. The {\tt R} and {\tt S-PLUS} function
{\tt smooth.spline()} uses 
$$\kappa_k\simeq \left({k+1}\over K+2\right)\mbox{th sample quantile of
the $x_i$'s}$$
where
\begin{displaymath}
K=\left\{
\begin{array}{ll}
n &  n< 50 \\
100& n=200 \\
140& n=800 \\
200+(n-3200)^{1/5} &     n>3200.\\
\end{array}
\right.
\end{displaymath}
Other values of $n$ between 50 and 3200 are
handled via a logarithmic interpolation. For many 
functional relationships, fewer knots are
sufficient. Figure \ref{fig:nbc} is one example, where
only $K=20$ interior knots are used without compromising
the quality of the fit. A common default in
the penalised spline literature is $K=\min(n_U/4,35)$,
where $n_U$ is the number of unique $x_i$'s
(e.g. Ruppert \etal, 2003). Ruppert (2002)
discusses `hi-tech' choice of $K$. 
The distribution of the knots, for a given $K$, may
have some affect on the results. As mentioned above,
{\tt smooth.spline()} uses quantile-based knots while
e.g. Eilers \& Marx (1996) recommend equally-spaced knots.
In most situations this effect will be minor. 
However, for either strategy, it is possible to
construct regression functions and  predictor variable 
distributions for which problems arise. More sophisticated
knot placement strategies may help. For example, 
Luo \& Wahba (1997) propose more sophisticated basis 
function reduction methods that could be adapted to the 
current context.

\section{Comparison with P-splines}\label{sec:comp}

The closest relatives of O'Sullivan penalised splines
are the P-splines of Eilers \& Marx (1996). If 
the interior knots $\kappa_5,\ldots,\kappa_{K+4}$ are
taken to be equally-spaced then the family of cubic
P-splines is given by (\ref{eq:fhat.osu}) with the 
$\bOmega$ replaced by $\bD_k\trans\bD_k$, where $\bD_k$ 
is the $k$th-order differencing matrix. This 
differencing penalty corresponds to a discrete approximation
to the integrated square of the $k$th derivative of
the B-spline smoother. 
The choice $k=2$ leads to the cubic P-spline estimate
\begin{equation}
\fhatP(x;\lambda)=\bB_x\bnuhatP,\quad\mbox{where}\quad
\bnuhatP\equiv(\bB\trans\bB+\lambda\bD_2\trans\bD_2)^{-1}\bB\trans\by,
\label{eq:fhat.psp}
\end{equation}
having the property that $\fhatP(\cdot;\lambda)$ approaches
the least squares line as $\lambda\to\infty$. In this sense,
(\ref{eq:fhat.psp}) is the closest relative of 
$\fhatO(\cdot;\lambda)$.
If the interior knots are
equally-spaced then the bands in the interior rows are,
up to multiplicative factors, as follows:

\begin{small}
\def\spcs{\ \ \ \ }
$$
\begin{array}{lrrrrrrr}
\mbox{O'Sullivan penalised splines (\ref{eq:fhat.osu}):}     
&\spcs   5,\spcs&    0,\spcs& -45,\spcs&  
80,\spcs& -45,\spcs&    0,\spcs&   5\\
\mbox{Cubic P-splines; 2nd order diff. (\ref{eq:fhat.psp}):} 
&\spcs  -4,\spcs&  24,\spcs& -60,\spcs&  
80,\spcs& -60,\spcs&  24,\spcs&  -4\\
\end{array}
$$
\end{small}

\begin{figure}[ht]
\null\centerline{\includegraphics[width=12cm]{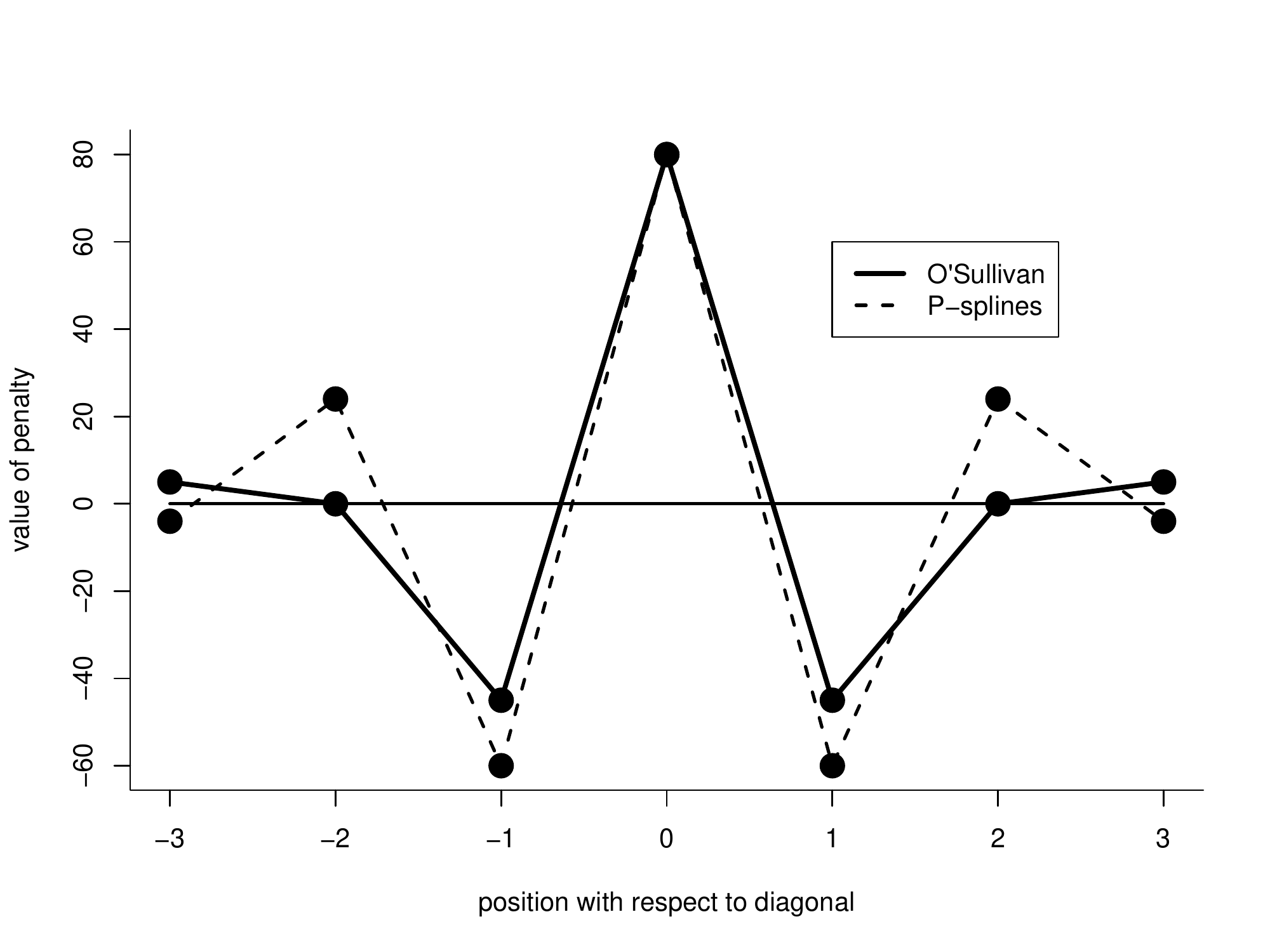}}
\caption
{\it
Comparison of near-diagonal entries of the penalty matrices
for O'Sullivan penalised splines and cubic P-splines with $k=2$
and equally-spaced interior knots.
}
\label{fig:penBspComp}
\end{figure}

\noindent
Figure \ref{fig:penBspComp} facilitates visual comparison of the two.
It is seen that the differences are relatively small, although not
negligible. 

What are the relative advantages of smoothers based on
cubic P-splines and O'Sullivan penalised splines, or O-splines
for short? Theoretical comparison between P-splines and O-splines
in terms of estimation performance, perhaps in the spirit 
of Hall \& Opsomer (2005), would be ideal -- although
beyond the scope of the current paper.

Eilers and Marx (1996) partially justify use of P-splines
rather than O'Sullivan splines based on simplicity of the
P-spline penalty matrix. However, as seen from (\ref{eq:OmegaExact}), 
the penalty matrix needed for O-splines can be obtained 
straightforwardly. Furthermore the discrete approximation
of P-splines requires equally-spaced knots which, depending on $f$,
may not be desirable, 

A possible advantage of P-splines is the option of 
higher-order penalties, although the resulting
smoothers can have erratic extrapolation behaviour.
A possible advantage 
of O-splines is their direct relationship 
with time-honoured smoothing splines, and their
attractive theoretical properties (e.g. Nussbaum, 1985).
>From the results described in Section \ref{sec:osu} is
clear that O-splines approach smoothing splines as
$K\to n$. But how close are O-splines to smoothing
splines for common (smaller) choices of $K$, and 
are they closer than P-splines with the same value
of $K$ and interior knots? To address these questions
we conducted an empirical study based on the
eighteen homoscedastic nonparametric 
regression settings in Wand (2000). For O-splines
we used $K=100$ equally spaced interior knots with 4 
repeated knots at each boundary as
described in Section \ref{sec:osu}. However, for
P-splines we used the knot sequence described
in the Appendix of Eilers \& Marx (1996) which
involves extending the knots beyond the boundary
rather than repeating them.
For each setting 200 samples were generated and 
smoothing spline estimates $\fhatS$, 
with smoothing parameter chosen via generalised 
cross-validation, were obtained. We then computed
$\fhatO$ and $\fhatP$ to have the same effective 
degrees of freedom as $\fhatS$ and recorded 
closeness measures $d(\fhatO,\fhatS;A)$ and $d(\fhatP,\fhatS;A)$
where
$$d(f,g;A)\equiv\int_A(f-g)^2.$$
We took $A$ corresponding to the intervals
$(a,\kappa_5)$ (left boundary), $(\kappa_5,\kappa_{K+5})$
(interior), $(\kappa_{K+5},b)$ (right boundary) and $(a,b)$
(total region) where the $\kappa_k$ denote the
knots used for the O-spline fits. The Wand (2000) settings all involve 
predictor data within the unit interval.
We took $(a,b)=(-0.1,1.1)$ to assess behaviour beyond the 
range of the data. 
Wilcoxon 
tests on the 200 differences $d(\fhatO,\fhatS;A)-d(\fhatP,\fhatS;A)$
were carried out
for each setting and choice of $A$. Apart from being distribution-free,
Wilcoxon tests have the advantage of being invariant to normalisation
and whether differences or ratios are used.
In all 72 cases O-splines were closer to smoothing splines
than P-splines in the sense that the Wilcoxon p-value$<0.01$.

%Figure \ref{fig:OSvsPS} summarises the results.
%\begin{figure}[ht]
%%\null%\null\centerline{\includegraphics[width=15cm]{OSvsPS.pdf}}
%\caption
%{\it
%Results of Wilcoxon tests measuring the relative closeness of
%O-splines and P-splines to smoothing splines for 
%the homoscedastic simulation settings of Wand (2000).
%The shading corresponds to significance at the 0.01 level.
%}
%\label{fig:OSvsPS}
%\end{figure}
%O-splines are seen to be
%statistically closer to smoothing splines than P-splines
%in 60 of the 72 situations, and all 36 of the boundary
%situations. It is clear that O-splines mimic the boundary 
%behaviour of smoothing splines more closely than P-splines.

\begin{figure}[ht]
\null\centerline{\includegraphics[width=15cm]{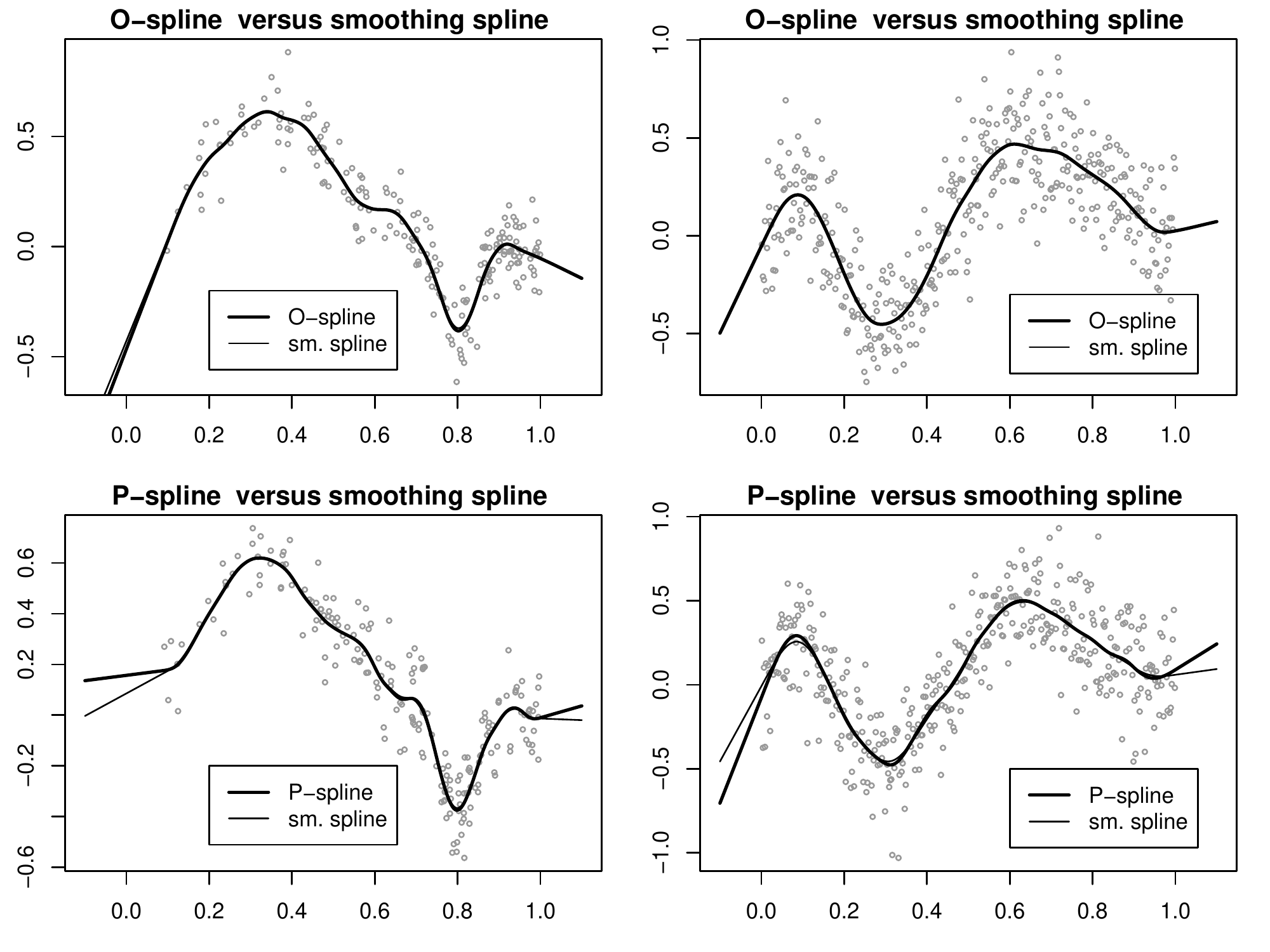}}
\caption
{\it
O-spline and P-spline fits compared with smoothing spline
fits corresponding to the 90th percentiles of the
$d(\fhatO,\fhatS;A)$ and $d(\fhatP,\fhatS;A)$ samples;
for two of the homoscedastic settings of Wand (2000).
}
\label{fig:specEx}
\end{figure}

To appreciate the practical significance of these 
results we plotted the data and estimates at the 90th percentiles
of each of the $d(\fhatO,\fhatS;A)$ and $d(\fhatP,\fhatS;A)$
samples, corresponding to relatively high discrepancies.
Some examples are shown in Figure \ref{fig:specEx}.

\noindent
In the interior all estimates of the same data, 
and same degrees of freedom, are almost indistinguishable
with the naked eye.
However, big differences occur at the boundary. P-splines
have a tendency to deviate from the natural boundary
behaviour of smoothing splines.
We also observed this phenomenom in the other 16 settings.
Further study into this differing extrapolation behaviour
would be worthwhile. We speculate that it comes
from differences between the exact integral penalty and its
discrete approximation near the boundary.

\section{Mixed Model Formulation}\label{sec:mix}

\begin{figure}[ht]
\null\centerline{\includegraphics[width=12cm]{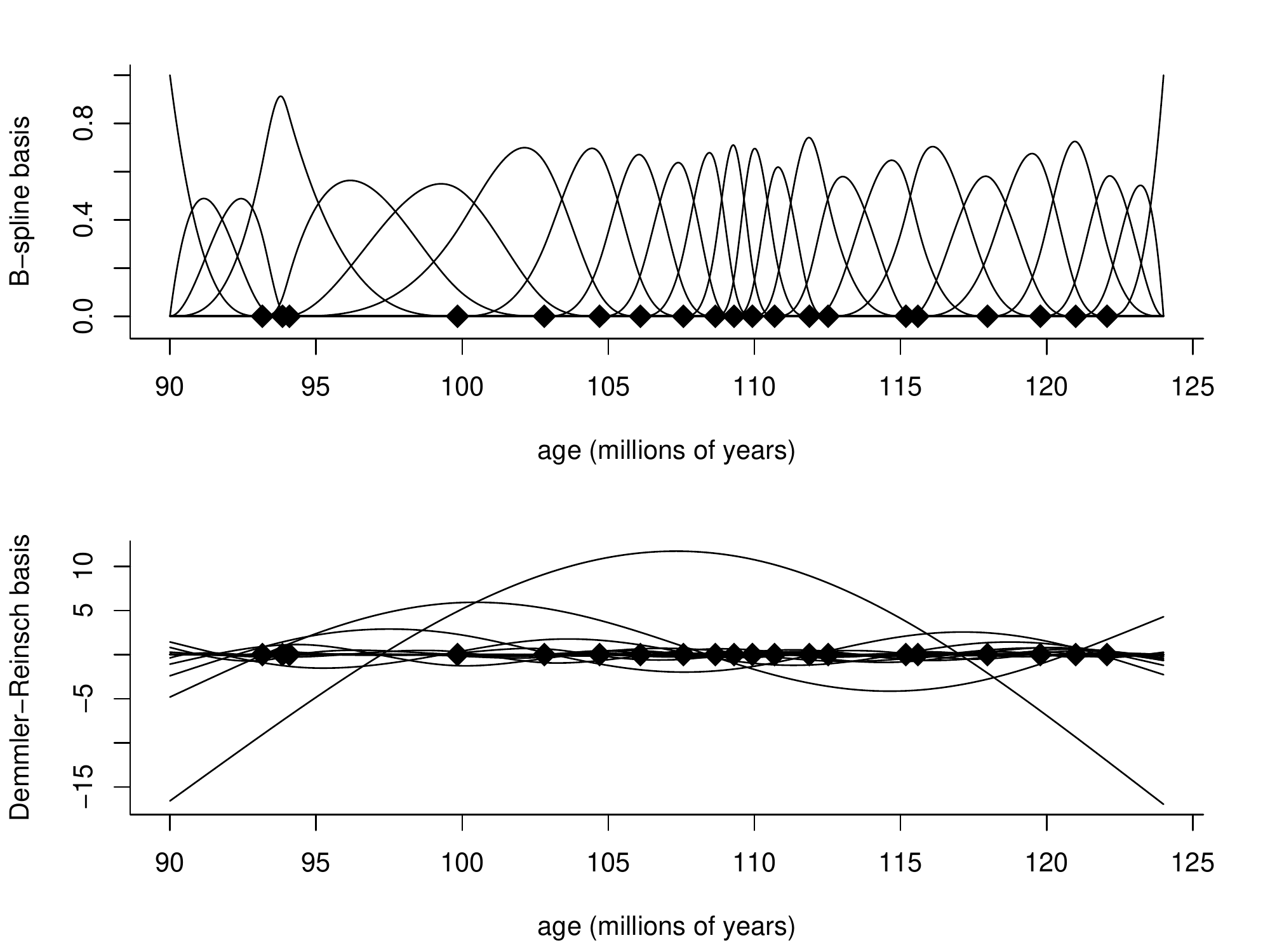}}
\caption
{\it
Comparison of B-spline basis and Demmler-Reinsch basis 
for the fossil data example of Figure \ref{fig:penBspComp}.
The interior knots are shown as solid diamonds ($\blacklozenge$).
}
\label{fig:basiscomp}
\end{figure}

There are several ways by which $\bnuhatO$ in (\ref{eq:fhat.osu})
can be expressed as a best linear unbiased predictor (BLUP) in a 
mixed model (e.g. Speed, 1990; Verbyla, 1994). However, from
a software standpoint, the most convenient form is 
$\bnuhatO=(\bbetahat,\buhat)$
%$\bnuhatO=[\bbetahat\trans \buhat\trans]\trans$
where $\bbetahat$ and $\buhat$ are (empirical) BLUPs of $\bbeta$
and $\bu$ in the mixed model
\begin{equation}
\by=\bX\bbeta+\bZ\bu+\bvarepsilon,\quad 
\left[
\begin{array}{c}
\bu\\
\bvarepsilon\\
\end{array}
\right]
\sim
N\left(\left[
\begin{array}{c}
\bzero\\
\bzero\\
\end{array}
\right],
\left[
\begin{array}{cc}
\sigma_u^2\bI & \bzero\\
\bzero & \sigma_{\varepsilon}^2\bI\\
\end{array}
\right]
\right)
\label{eq:mixmod}
\end{equation}
for some design matrices $\bX$ and $\bZ$. 
An explicit expression for the BLUP in (\ref{eq:mixmod})
(e.g. Ruppert, Wand \& Carroll, 2003; Section 4.5.3) is 
$$\left[\begin{array}{c}
\bbetahat\\
\buhat\\
\end{array}\right]=
\bnuhatO=\left(\bC\trans\bC+\lambda\left[\begin{array}{cc}
\bzero\ \ \ &\ \ \ \bzero\\
\bzero\ \ \ &\ \ \ \bI\\
\end{array}
\right]\right)^{-1}
\bC\trans\by,\quad \lambda=\sigma_u^2/\sigma^2_{\varepsilon},$$
where $\bC=[\bX|\bZ]$, 
$\bI$ is the identity matrix with the same number of columns
as $\bZ$. This `canonical form' can be achieved if a 
$(K+4)\times(K+4)$ linear transformation
matrix $\bL$ can be found such that $\bC=\bB\bL$ and 
$$\bL\trans\bOmega\bL=\left[\begin{array}{cc}
                         \bzero\ \ \  &\ \ \ \bzero\\
                         \bzero\ \ \  &\ \ \  \bI\\\end{array}\right].
$$
The usual method for obtaining $\bL$ is spectral decomposition
(e.g. Nychka \& Cummins, 1996; Cantoni \& Hastie, 2002;
Welham \etal, 2007).
It follows from results in the smoothing spline literature 
(e.g. Speed, 1991, Section 6) that 
$$\mbox{rank}(\bOmega)=K+2.$$
Hence, the spectral decomposition of $\bOmega$ is of
the form $\bOmega=\bU\diag(\bd)\bU\trans$ where $\bU\trans\bU=\bI$
and $\bd$ is a $(K+4)\times 1$ vector with exactly 2 zero entries
and $K+2$ positive entries. Let $\bdZ$ be the $(K+2)\times 1$ 
sub-vector of $\bd$ containing these positive entries, and let 
$\bUZ$ be the $(K+4)\times(K+2)$ sub-matrix of $\bU$
with columns corresponding to positive entries of $\bd$. 
Then an appropriate linear transformation is 
$\bL=[\bUX|\bUZ\diag(\bdZ^{-1/2})]$.
This leads to the fixed and random effects design matrices:
\begin{equation}
\bX=\bB\bUX\quad\mbox{and}\quad\bZ=\bB\bUZ\diag(\bdZ^{-1/2}).
\label{eq:XZ}
\end{equation}
However, following again from the aforementioned smoothing
spline literature (e.g. Speed, 1991, Section 6),
$\bB\bUX$ is a basis for the space of 
straight lines so the simpler specification 
$\bX=[1\ x_i]_{1\le i\le n}$
may be used instead without affecting the fit. The
spline basis formed through spectral decomposition of 
more familiar bases is known as the {\em Demmler-Reinsch}
basis (e.g. Nychka \& Cummins, 1996).
Figure \ref{fig:basiscomp} allows comparison of the original
B-spline basis, corresponding to $\bB$, 
and the Demmler-Reinsch basis corresponding to $\bZ$.
Notice the damping of the $\bZ$ basis functions with increasing
oscillation. This compensates for the fact that the 
penalty is a multiple of the identity matrix.

In the Appendix it is shown how the {\tt R} linear mixed model function
{\tt lme()} can be used to obtain $\fhatO(\cdot;\lambda)$ based
on (\ref{eq:mixmod}), with $\bZ$ given by (\ref{eq:XZ}).
For simple scatterplot smoothing there is little difference
between this approach and direct use of {\tt smooth.spline()},
and the answers are equivalent if the knot sequence and 
$\lambda$ values are equal. The default choice of $\lambda$
differs: {\tt lme()} uses restricted maximum likelihood (REML)
to choose $\lambda$, while {\tt smooth.spline()} uses generalised
cross-validation (GCV).
 
\begin{figure}[ht]
\null\centerline{\includegraphics[width=14cm]{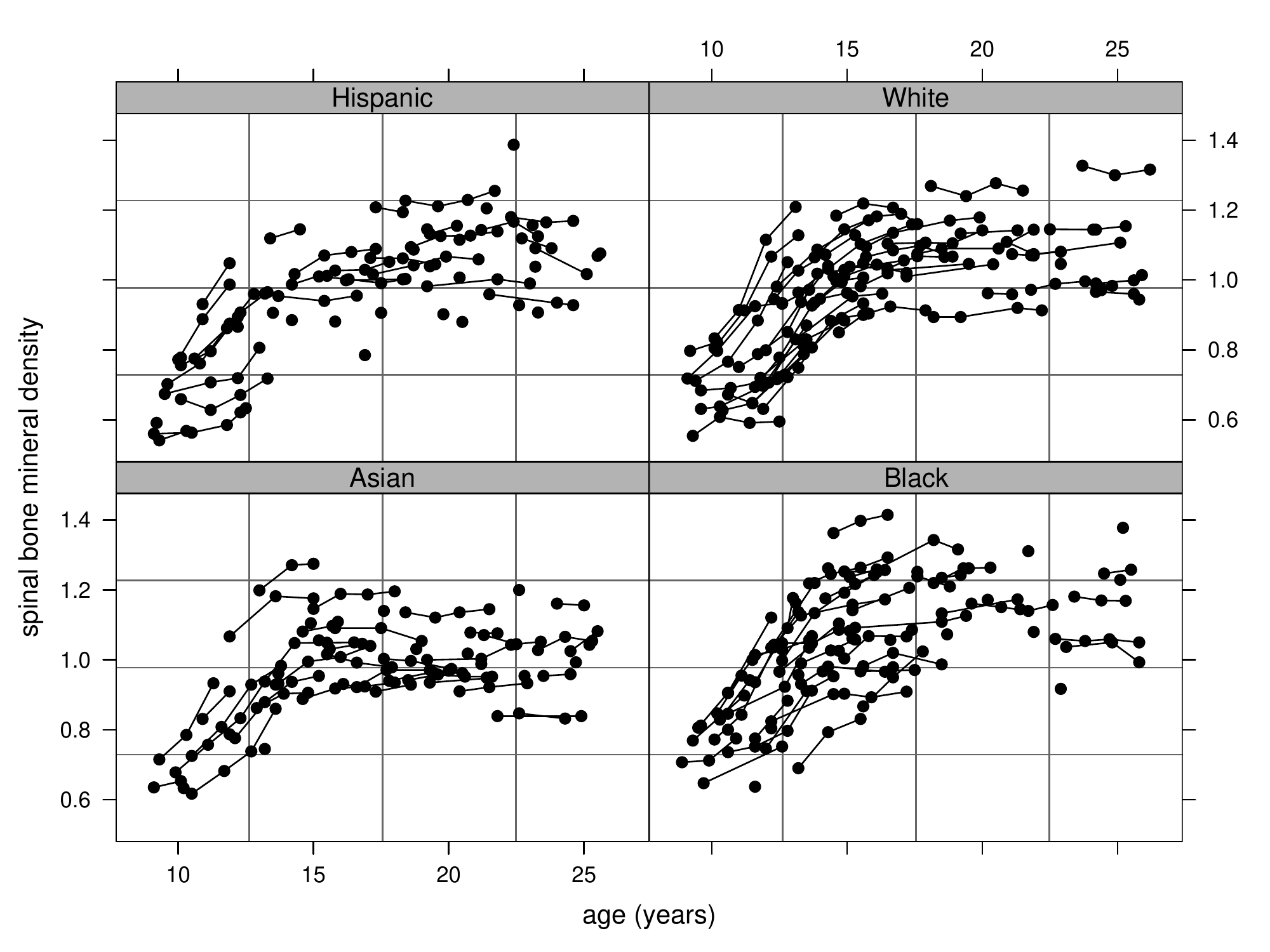}}
\caption
{\it The spinal bone mineral data. Lines connect measurements
taken on the same subject.}
\label{fig:sbmdRaw}
\end{figure}

The main advantage of the mixed model formulation of penalised splines
is the incorporation into more complex models. Several examples
are given in, for example, Ruppert, Wand \& Carroll (2003). 
We will briefly describe one of them here.
Figure \ref{fig:sbmdRaw} displays a longitudinal data set
on bone mineral acquisition in young females
(source: Bachrach, Hastie, Wang, Narasimhan \& Marcus, 1999).
The data consists of spinal bone mineral density (SBMD) 
measurements on each of 230 female subjects aged between
8 and 27. Each subject is measured between one and four times.
Let $n_i$ denote the number of measurements for subject $i$.
The subjects have been divided into four ethnic groups:
Asian, Black, Hispanic and White.

A useful additive mixed model for these data is:
\begin{equation}
\SBMD_{ij}=U_i+f({\tt age}_{ij})+\beta_1{\tt Black_i}
+\beta_2{\tt Hispanic_i}+\beta_3{\tt White_i}
+\varepsilon_{ij},\quad
1\le i\le 230,\quad 1\le j\le n_i\quad
\label{eq:sbmd.model}
\end{equation}
where $U_i\ \mbox{i.i.d.}\ N(0,\sigu^2)$  
are random intercepts for each subject,
and ${\tt Black}_i$, ${\tt Hispanic}_i$ and 
${\tt White}_i$ are ethnicity indicators.
$\varepsilon_{ij}\ \mbox{i.i.d.}\ N(0,\sigeps^2)$ are random errors.
More sophisticated models that account for, say, serial
correlation could be entertained. 

O'Sullivan penalised splines can be used to 
fit (\ref{eq:sbmd.model}) with the 
design matrices set up as follows.
Based on the ${\tt age}_i$ values and appropriate knots, 
set up 
$$\bZspline=\bB\bUZ\diag(\bdZ^{-1/2})$$
analogous to the $\bZ$ matrix of (\ref{eq:XZ}) for simple
scatterplot smoothing. In the Appendix, when fitting
data of this type, we use 15 interior
knots corresponding to quantiles of the unique age values.
Form
$$\bX=\left[
\begin{array}{ccccc}
1     &{\tt age}_{11}     &  {\tt Black}_1      
& {\tt Hispanic}_1 & {\tt White}_1   \\
\vdots&\vdots    &   \vdots            &  \vdots          & \vdots \\
1     &{\tt age}_{1n_1}    &  {\tt Black}_1      
& {\tt Hispanic}_1 & {\tt White}_1 \\
\vdots&\vdots    &   \vdots            &  \vdots         & \vdots\\
1     &{\tt age}_{230,1}    &  {\tt Black}_{230}  
& {\tt Hispanic}_{230}& {\tt White}_{230}    \\   
\vdots&\vdots    &   \vdots            &  \vdots         & \vdots\\
1     &{\tt age}_{230,n_{230}} &  {\tt Black}_{230}  
& {\tt Hispanic}_{230}& {\tt White}_{230}   \\
\end{array}
\right]
\ \mbox{and}\ 
\bZsubj=\left[
\begin{array}{cccc}
1         &     0      & \cdots    &  0    \\
\vdots    &   \vdots   & \cdots   &  \vdots         \\
1         &     0      & \cdots    & 0 \\
\vdots    &   \vdots   & \ddots    &  \vdots         \\
0         &   0        & \cdots   & 1   \\
\vdots    &   \vdots   & \cdots    &  \vdots         \\
0         &   0        & \cdots    & 1 \\
\end{array}
\right].
$$
Concatenate $\bZsubj$ and $\bZspline$ to form 
$$\bZ=[\bZsubj|\bZspline].$$
The appropriate mixed model is then
\begin{equation}
\by = \bX \bbeta + \bZ \bu + \beps , \quad
\Cov\left[ \begin{array}{c} \bu \\ \beps \end{array} \right]
        =\left[\begin{array}{ccc}
                \sigma_U^2\bI & \bzero & \bzero \\
                \bzero & \sigma_u^2\bI & \bzero \\
                \bzero & \bzero        & \sigma^2_{\varepsilon}\bI \\ 
               \end{array}
          \right].
\label{eq:sbmd.lmm}
\end{equation}

The Appendix contains {\tt R} code for fitting this model. 
Note, in particular, that it circumvents 
explicit specification of $\bZsubj$. This is important
for large longitudinal datasets.

\section{Bayesian Analysis and Markov Chain Monte Carlo}\label{sec:baye}

A particularly attractive advantage of penalised splines, compared
with smoothing splines, is the ease with which they can be fed
into Markov Chain Monte Carlo (MCMC) schemes for fitting Bayesian
semiparametric regression models -- due to the reduction in
the number of basis functions.
For simple scatterplot smoothing
this involves the Bayesian version of (\ref{eq:mixmod}):
$$
\by|\bbeta,\bu,\sigma_{\varepsilon}^2\sim 
N(\bX\bbeta+\bZ\bu,\sigma_{\varepsilon}^2\bI),\quad
\bu|\sigma_u^2\sim N(\bzero,\sigma_u^2\bI)
$$
and suitable (usually diffuse) prior distributions for 
$\bbeta$, $\sigma_u^2$ and $\sigma_{\varepsilon}^2$.
However, the big advantages of a Bayesian/MCMC
approach are realised when handling complications such as 
measurement error (e.g. Carroll, Ruppert, 
Stefanski \& Crainiceanu, 2006) and
generalised responses (e.g. Zhao, Staudenmayer, Coull \& Wand, 2006),
which are hindered by intractable integrals in the likelihood.

Crainiceanu, Ruppert \& Wand (2005) focus on use of the 
MCMC package {\tt WinBUGS} 
({\tt Windows} version of {\tt BUGS}, Spiegelhalter, Thomas \& Best, 2000)
for Bayesian penalised spline models.
They reported that the choice of basis functions can have a 
substantial impact on the convergence of the chain. We decided
to conduct some convergence checks for MCMC
fitting of the regression model:
\begin{equation}
\logit\{P(\mbox{\tt union}_i=1|\mbox{\tt wage}_i)\}
=f(\mbox{\tt wage}_i)
\label{eq:tumodel}
\end{equation}
with $f$ estimated via O'Sullivan penalised splines.
Here $({\tt wage}_i,{\tt union}_i)$, $1\le i\le 534$, are 
pairs of wage amounts (dollars per hour) and trade union membership
indicators for a sample of U.S. workers (source: Berndt, 1991).
We expressed (\ref{eq:tumodel}) as the Bayesian logistic mixed model:
$$\logit\{P(\mbox{\tt union}_i=1|\mbox{\tt wage}_i)\}
=(\bX\bbeta+\bZ\bu)_i,\quad 1\le i\le 534
$$
where  $\bX=[1\ {\tt wage}_i]_{1\le i\le 534}$ and
$\bZ=\bB\bUZ\diag(\bdZ^{-1/2})$, using the notation 
of Section \ref{sec:mix}. We used 15 interior knots
with quantile spacing.

\begin{figure}[h]
\null\centerline{\includegraphics[width=10cm]{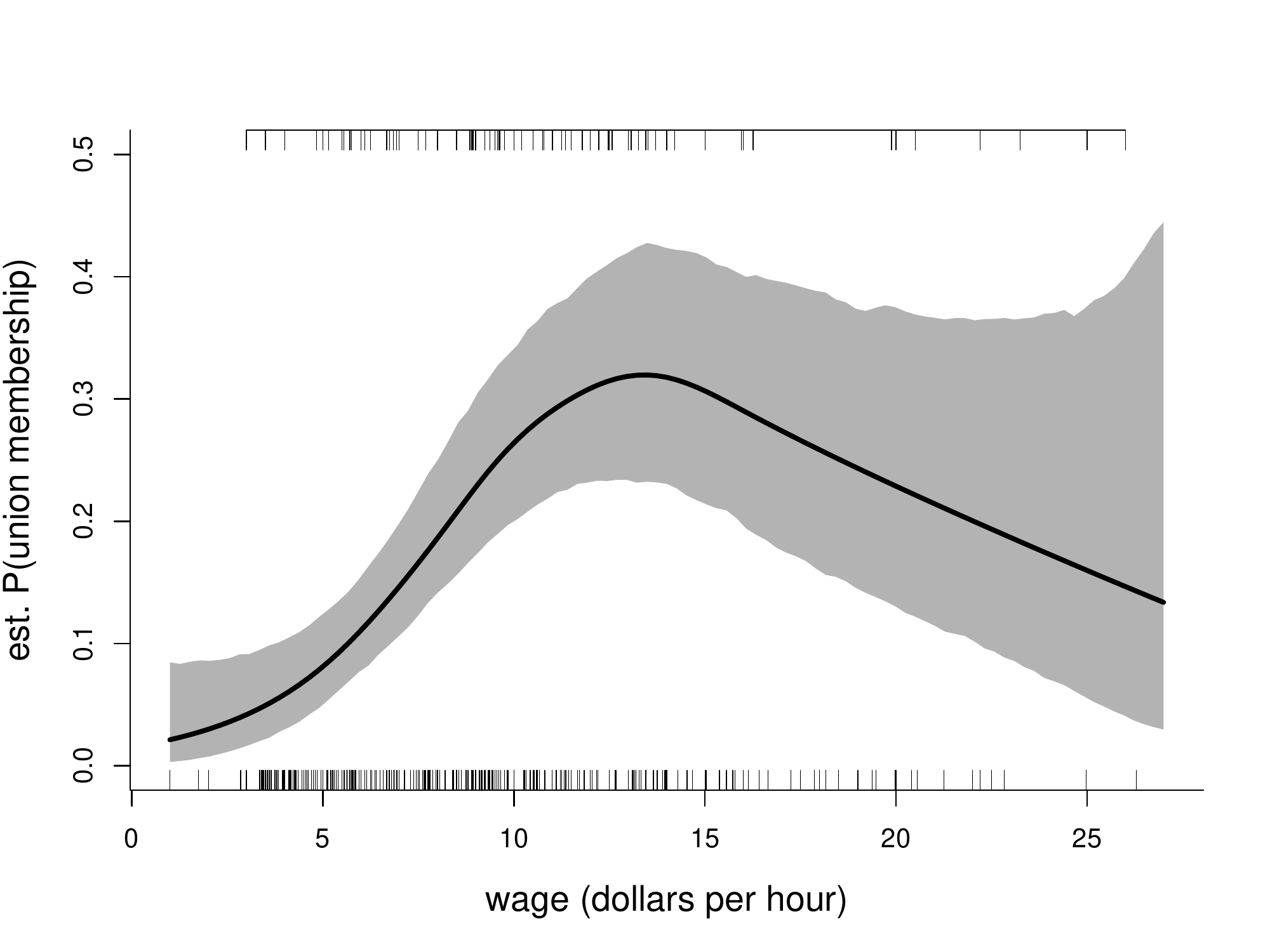}}
\caption
{\it
Fit of (\ref{eq:tumodel}) using O'Sullivan penalised splines.
}
\label{fig:tuOSfit}
\end{figure}

\begin{figure}[!h]
\null\centerline{\includegraphics[width=\textwidth]{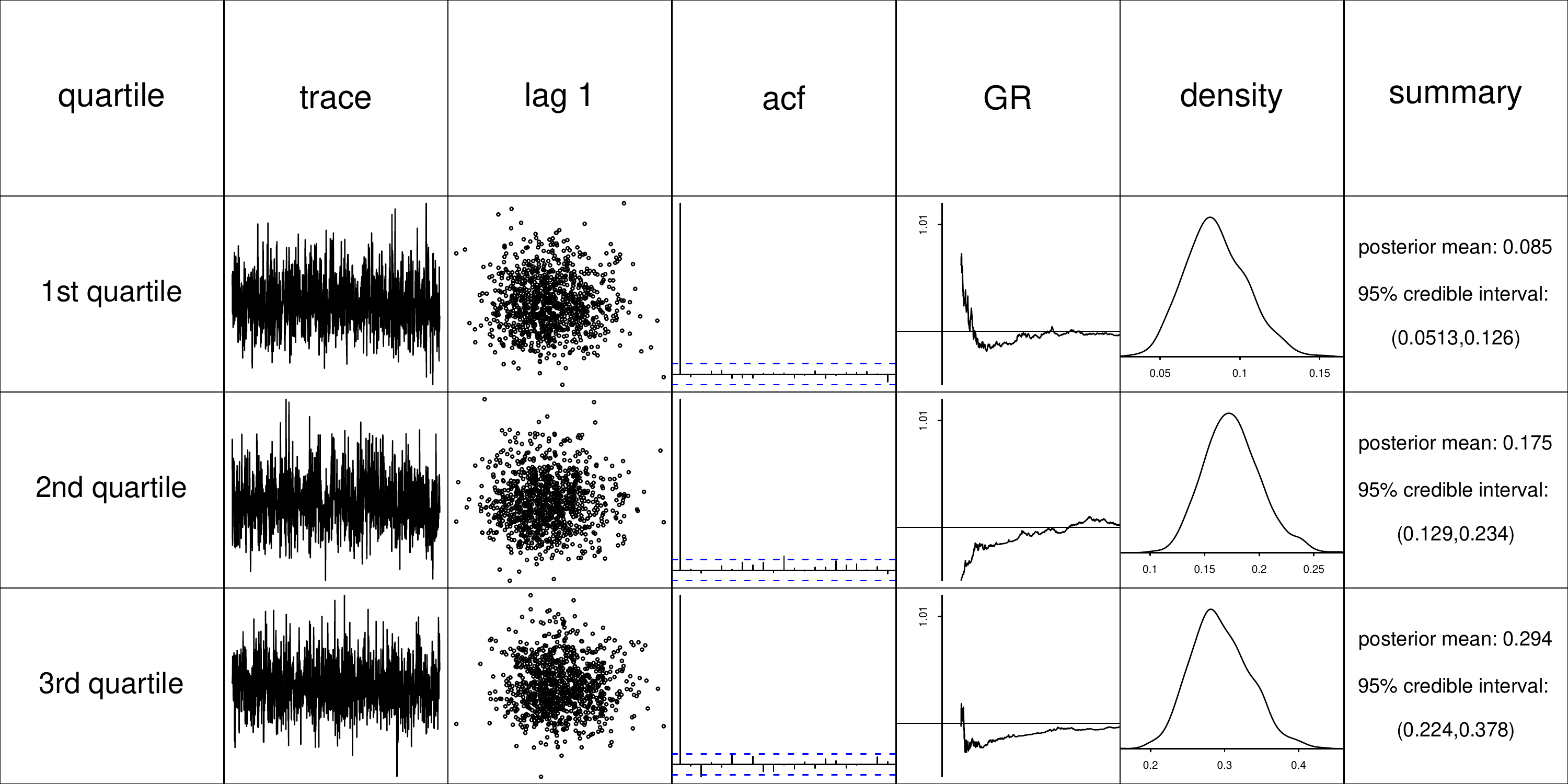}}
\caption
{\it
Assessment of MCMC convergence 
for O'Sullivan penalised spline estimation of 
(\ref{eq:tumodel})  at each quartile of wage. 
The columns are: quartile of wage,
trace plot of sample of corresponding coefficient, plot 
of sample against 1-lagged sample, sample autocorrelation
function, Gelman-Rubin $\surd{\Rhat}$ diagnostic,
kernel estimates posterior density and 
basic numerical summaries. }
\label{fig:mixOS}
\end{figure}

Following the advice of Zhao \etal\ (2006) we used 
{\tt WinBUGS} to generate chains
of length 5000 after a burn-in of 5000 and applied a thinning
factor of 5, resulting in posterior samples of size 1000.
Also in keeping with the recommendations of  Zhao \etal\ (2006)
we placed diffuse priors on the fixed effect parameters
and variance component: $\beta_0,\beta_1$ independent $N(0,10^8)$
and the prior density of $\sigma_u^2$ proportional to 
$(\sigma_u^2)^{-1.01}e^{-1/(100\sigma_u^2)}$, 
the Inverse Gamma distribution with shape and rate parameter
both 0.01, after scaling the predictor to have unit variance.
Zhao \etal\ (2006) found that the results can be sensitive
to the choice of the Inverse Gamma hyperparameter.

The pointwise posterior mean effect of {\tt wage} on the
probability of trade union membership, together with 95\% pointwise 
credible sets, is shown in Figure \ref{fig:tuOSfit}.
Figure \ref{fig:mixOS} allows assessment of convergence of
the MCMC at each quartile of the wage sample and is 
seen to be excellent in each case.
We also conducted convergence checks for larger logistic additive
models involving up to 6 predictors and 3 smooth functions and
found the mixing to be very good when O-splines were used.

Several examples of semiparametric regression with
{\tt WinBUGS}, including code, are given in
Crainiceanu \etal (2005) and Zhao \etal (2006).

\section{General Degree Extension}\label{sec:hdOsul}

Cubic O'Sullivan penalised splines have a natural extension to
general odd degree splines. Higher degree splines have a role
to play when smoother curve estimates are required. This
arises, for example, in feature significance methodology
(e.g. Chaudhuri \& Marron, 1999; Hannig \& Marron, 2006)
where first and second derivatives of the fit are required.

Return to the simple nonparametric regression setting 
(\ref{eq:simpNR}) and let $m$ be a general positive integer. 
Form the knot sequence
$$a= \kappa_1 = \ldots = \kappa_{2m} < \kappa_{2m+1} < \ldots
< \kappa_{2m+K} < \kappa_{2m+K+1} = \ldots = \kappa_{4m+K}=b$$
and let $B_{2m-1,1},\ldots,B_{2m-1,K+2m}$ be the degree $(2m-1)$
B-spline basis defined by these knots.
Order $m$ O'Sullivan penalised splines then take the general form
$$
\fhatO(x;m,\lambda)\equiv\bB_{2m-1,x}\bnuhatO\quad\mbox{where}\quad
\bnuhatO\equiv(\bB_{2m-1}\trans\bB_{2m-1}+
\lambda\bOmega^{(m)})^{-1}\bB_{2m-1}\trans\by.
$$
Here $\bB_{2m-1}$ is the $n\times(K+4m)$ design matrix with $(i,k)$ entry
$B_{{2m-1},k}(x_i)$, $\bB_{2m-1,x}=[B_{2m-1,1}(x),\ldots,B_{2m-1,K+2m}(x)]$ 
and $\bOmega^{(m)}$ is the $(K+2m)\times(K+2m)$ 
penalty matrix with $(k,k')$ entry
$$\bOmega^{(m)}_{kk'}=\int_a^b B_{2m-1,k}^{(m)}(x)B_{2m-1,k'}^{(m)}(x)\,dx.$$
In the special case where the interior knots coincide with the $x_i$'s,
assumed distinct, $\fhatO(\cdot;m,\lambda)$ corresponds to the order $m$
smoothing spline; i.e. the minimiser of 
$$\sum_{i=1}^n\{y_i-f(x_i)\}^2 + \lambda\int_a^b f^{(m)}(x)^2\,dx$$
(e.g. Schoenberg, 1964).

We are now ready to state our result for exact computation of O'Sullivan
spline penalty matrices:

\jump
\noindent
{\bf Theorem.}
\begin{sl} The penalty matrix
$\bOmega^{(m)}$ admits the exact explicit expression
$$\bOmega^{(m)}=(\bBtilde^{(m)})^T\mbox{\rm diag}(\bw)\bBtilde^{(m)}$$
where $\bBtilde^{(m)}$ is the $(2m-1)(K+4m-1)\times(K+2m)$ 
matrix with $(i,j)$ entry
$B_{2m-1,j}^{(m)}(\xtilde_i)$ and $\bw$ is a 
$(2m-1)(K+4m-1)\times1$ vector with $i$th
entry $w_i$. The $\xtilde_i$ and $w_i$ values are obtained according to
$$\xtilde_{(2m-1)(\ell-1)+\ell'+1}=\kappa_{\ell}+\ell'h_{m,\ell},\quad
w_{(2m-1)(\ell-1)+\ell'+1}=h_{m,\ell}\omega_{m,\ell'}$$
for $1\le\ell\le K+4m-1$, $0\le \ell'\le 2m-2$. 
Here, for $1\le k\le K+2m$,
$h_{1,k}=\kappa_{k+1}-\kappa_{k}$ and, for $m\ge2$, 
$h_{m,k}=(\kappa_{k+1}-\kappa_{k})/(2m-2)$.
Lastly, for all $m\ge1$,
$$\omega_{m,k}=\frac{(-1)^k}{k!(2m-2-k)!}\int_0^{2m-2}
\frac{t(t-1)\cdots(t-2m+2)}{t-k}\,dt,
\quad k=0,\ldots,{2m-2}.$$
\end{sl}

\noindent
{\bf Proof.} The $(k,k')$ entry of $\bOmega^{(m)}$ is
\begin{equation}
\bOmega^{(m)}_{kk'}=
\int_a^b B_{2m-1,k}^{(m)}(x) B_{2m-1,k'}^{(m)}(x)\,dx=
\sum_{i=1}^{K+4m-1}\int_{\kappa_i}^{\kappa_{i+1}}
B_{2m-1,k}^{(m)}(x)B_{2m-1,k'}^{(m)}(x)\,dx.
\label{eq:int.breakdown}
\end{equation}
Since $B^{(m)}_{2m-1,k}(x),B^{(m)}_{2m-1,k'}(x)$ are 
degree $m-1$ polynomials on
each interval $x\in (\kappa_i,\kappa_{i+1})$ for 
$1\le i \le K+4m-1$ the function
$B_{2m-1,k}^{(m)}(x) B_{2m-1,k'}^{(m)}(x)$ is a degree $2(m-1)$ polynomial on
the same interval. The result follows by applying the Newton-Cotes integration
$(2m-1)$-point rule (e.g. Whittaker \& Robinson, 1967) to the 
right hand side of (\ref{eq:int.breakdown}) which is exact for polynomials of 
degree $2(m-1)$ or lower.

\begin{flushright}$\square$\end{flushright}

Table \ref{tab:omega.value} provides values of $\omega_{m,k}$
for O'Sullivan polynomials up to degree 7. This, together
with the Theorem, allows direct computation of penalty
matrices of O'Sullivan splines for $m\le 4$. Higher values
of $m$ require one-off calculation of the $\omega_{k,m}$ through, say,
a symbolic computation package such as {\tt Maple}.

\begin{table}[ht]
\begin{center}
\begin{tabular}{c|ccccccc}
\hline
$\quad m$$\backslash$$k$ & 0 & 1 & 2 & 3 & 4 & 5 & 6 \\
\hline
1                & 1 &   &   &   &   &   &   \\ 
2                &1/3&4/3&1/3&   &   &   &   \\ 
3                &14/45&64/45&8/15&64/45 & 14/45 &  & \\
4                &41/140&54/35&27/140&68/35&27/140&54/35&41/140\\
\hline
\end{tabular}
\end{center}
\caption{\it Table of $\omega_{m,k}$ values for $m\le 4$.}
\label{tab:omega.value}
\end{table}

Recall from Section \ref{sec:osu} that, in the case of 
cubic O'Sullivan splines,
Newton-Cotes integration reduces to Simpson's rule and a simpler, 
more revealing,
expression results in the shape of (\ref{eq:OmegaExact}).

\section{Closing Remarks}\label{sec:discn}

Smoothing splines have a special place in 
nonparametric and semiparametric
regression. They are based on simple and intuitive principles,
have an attractive theory (e.g. Nussbaum, 1985; Wahba, 1990;
Eubank, 1994 and Solo, 2000) and possess 
good practical  properties such as natural boundary behaviour. 
Penalised splines, including P-splines, have gained
popularity for reasons stated in the Introduction.
However, proponents of penalised splines
have been viewed by some, especially 
in the smoothing spline community, as ignoring
the benefits of that have been established
for smoothing splines over the past few
decades. O'Sullivan penalised splines, being
a direct generalisation and closer
approximation of smoothing splines,
provide an attractive link between the two streams
of semiparametric regression research and
allow analysts to enjoy the best of both worlds. 

\section*{Acknowledgements}

We are grateful for comments from an associate
editor and a referee, that led to considerable
improvements in this article. The first author
would also like to acknowledge the hospitality
of the Indian Statistical Institute, Calcutta,
where some of this research took place.

\section*{Appendix: R implementation}

In this Appendix we provide {\tt R} code for use of
O'Sullivan penalised splines in the simplest semiparametric
regression setting: scatterplot smoothing. 
The extensions to more complex models, such as those described
by Ngo \& Wand (2004) and Crainiceanu, Ruppert \& Wand (2005),
is straightforward. We illustrate one of these
extensions: additive mixed models.

\jump
\leftline{\underline{\sl Direct scatterplot smoothing  with  
user choice of smoothing parameter}}
\jump

\noindent
Obtain scatterplot data corresponding to environmental
data from the {\tt R} package {\tt lattice}. Set up plotting 
grid, knots and smoothing parameter:

\begin{small}
\begin{verbatim}
library(lattice) ; attach(environmental)
x <- radiation ; y <- ozone^(1/3)
a <- 0 ; b <- 350 ; xg <- seq(a,b,length=101)
numIntKnots <- 20 ; lambda <-  1000
\end{verbatim}
\end{small}
   
\noindent
Set up the design matrix and related quantities:

\begin{small}
\begin{verbatim}
library(splines)
intKnots <- quantile(unique(x),seq(0,1,length=
        (numIntKnots+2))[-c(1,(numIntKnots+2))])
names(intKnots) <- NULL
B <- bs(x,knots=intKnots,degree=3,
      Boundary.knots=c(a,b),intercept=TRUE)
BTB <- crossprod(B,B) ; BTy <- crossprod(B,y)
\end{verbatim}
\end{small}
  
\noindent
Create the $\bOmega$ matrix:

\begin{small}
\begin{verbatim}
formOmega <- function(a,b,intKnots)
{
   allKnots <- c(rep(a,4),intKnots,rep(b,4)) 
   K <- length(intKnots) ; L <- 3*(K+8)
   xtilde <- (rep(allKnots,each=3)[-c(1,(L-1),L)]+ 
              rep(allKnots,each=3)[-c(1,2,L)])/2
   wts <- rep(diff(allKnots),each=3)*rep(c(1,4,1)/6,K+7)
   Bdd <- spline.des(allKnots,xtilde,derivs=rep(2,length(xtilde)),
                  outer.ok=TRUE)$design  
   Omega     <- t(Bdd*wts)%*%Bdd     
   return(Omega)
}

Omega <- formOmega(a,b,intKnots)   
\end{verbatim}
\end{small}    
     
\noindent
Obtain the coefficients:
        
\noindent
\begin{small}
\begin{verbatim}
nuHat <- solve(BTB+lambda*Omega,BTy)
\end{verbatim}
\end{small}

\noindent
For large $K$ the following alternative Cholesky-based
approach can be considerably faster ($O(K)$, because 
$\bB\trans\bB+\lambda\bOmega$ is banded diagonal):

\noindent
\begin{small}
\begin{verbatim}
cholFac <- chol(BTB+lambda*Omega)
nuHat <- backsolve(cholFac,forwardsolve(t(cholFac),BTy))  
\end{verbatim}
\end{small}    
      
\noindent
Display the fit:
         
\begin{small}
\begin{verbatim}
Bg <- bs(xg,knots=intKnots,degree=3,Boundary.knots=c(a,b),intercept=TRUE)
fhatg <- Bg%*%nuHat
plot(x,y,xlim=range(xg),bty="l",type="n",xlab="radiation",
     ylab="cuberoot of ozone",main="(a) direct fit; user 
     choice of smooth. par.")
lines(xg,fhatg,lwd=2)   
points(x,y,lwd=2)
\end{verbatim}
\end{small}   
 
\leftline{\underline{\sl Mixed model scatterplot smoothing 
with REML choice of smoothing parameter}}
\jump

\noindent
Obtain the spectral decomposition of $\bOmega$:

\begin{small}
\begin{verbatim}
eigOmega <- eigen(Omega)
\end{verbatim}
\end{small}
    
\noindent
Obtain the matrix for linear transformation of $\bB$ to $\bZ$:

\begin{small}
\begin{verbatim}
indsZ <- 1:(numIntKnots+2)
UZ <- eigOmega$vectors[,indsZ]
LZ <- t(t(UZ)/sqrt(eigOmega$values[indsZ]))
\end{verbatim}
\end{small}
            
\noindent
Perform stability check:

\begin{small}
\begin{verbatim}
indsX <- (numIntKnots+3):(numIntKnots+4)
UX <- eigOmega$vectors[,indsX]   
L <- cbind(UX,LZ)
stabCheck <- t(crossprod(L,t(crossprod(L,Omega))))          
if (sum(stabCheck^2) > 1.0001*(numIntKnots+2))
    print("WARNING: NUMERICAL INSTABILITY ARISING FROM SPECTRAL DECOMPOSITION")
\end{verbatim}
\end{small}
   
\noindent
Form the $\bX$ and $\bZ$ matrices:

\begin{small}
\begin{verbatim}
X <- cbind(rep(1,length(x)),x)
Z <- B%*%LZ
\end{verbatim}
\end{small}   
   
\noindent
Fit using {\tt lme()} with REML choice of smoothing parameter:

\begin{small}
\begin{verbatim}
library(nlme)
group <- rep(1,length(x))
gpData <- groupedData(y~x|group,data=data.frame(x,y))
fit <- lme(y~-1+X,random=pdIdent(~-1+Z),data=gpData)
\end{verbatim}
\end{small}   
   
\noindent
Extract coefficients and plot scatterplot smooth over a grid:

\begin{small}
\begin{verbatim}
betaHat <- fit$coef$fixed
uHat <- unlist(fit$coef$random)
Zg <- Bg%*%LZ
fhatgREML <- betaHat[1] + betaHat[2]*xg + Zg%*%uHat
plot(x,y,xlim=range(xg),bty="l",type="n",xlab="radiation",
      ylab="cuberoot of ozone",main="(b) mixed model fit; 
      REML choice of smooth. par.")
lines(xg,fhatgREML,lwd=2)   
points(x,y,lwd=2)
\end{verbatim}
\end{small}   

\begin{figure}[ht]
\null\centerline{\includegraphics[width=12cm]{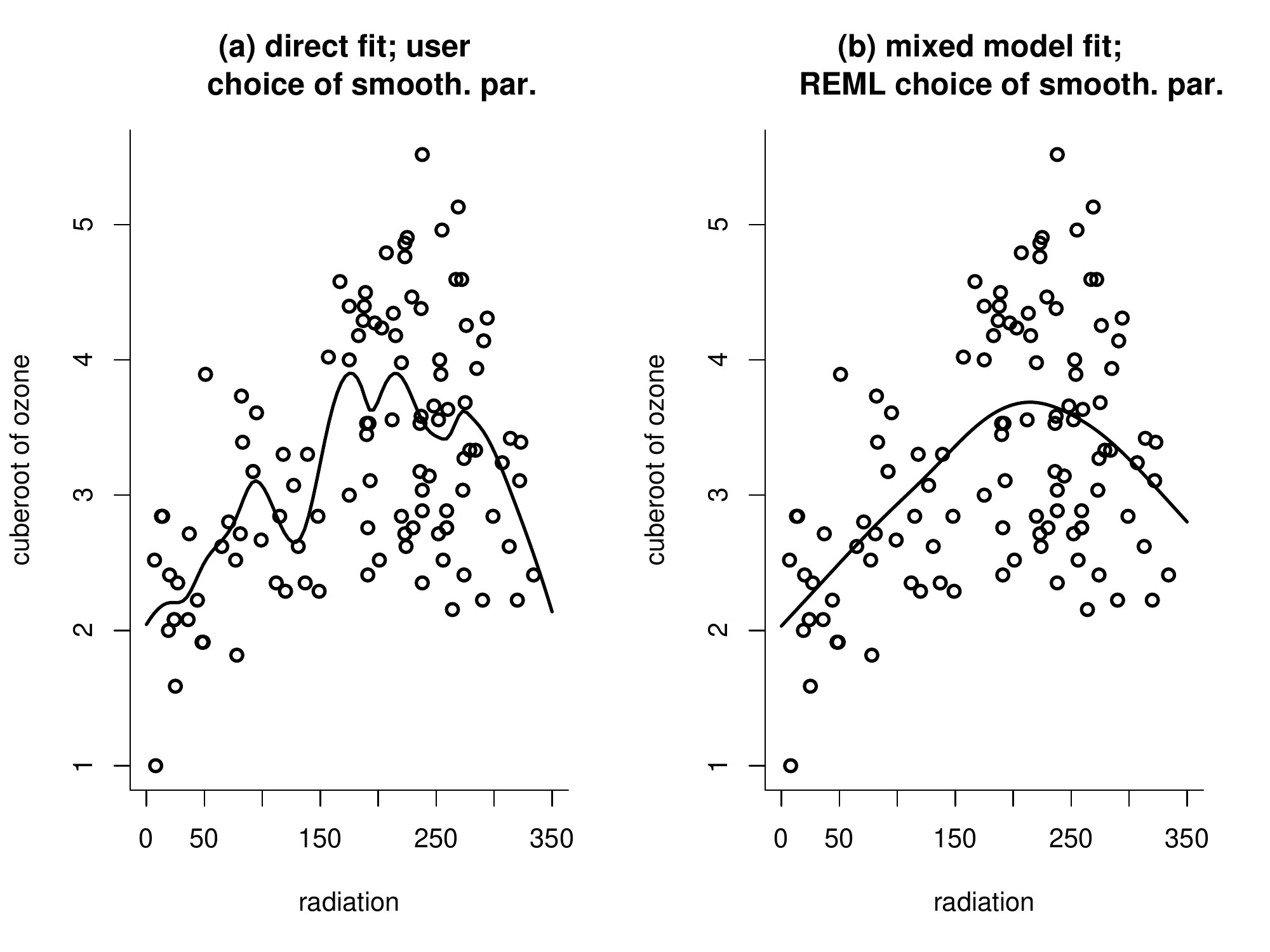}}
\caption
{\it
   Plots obtained from execution of the first two chunks
   of code in this Appendix.
}
\label{fig:osuScattSm}
\end{figure}

\noindent
Execution of the above code leads to Figure \ref{fig:osuScattSm}.

\jump\jump
\leftline{\underline{\sl Fitting an additive mixed model}}
\jump

\noindent
 The spinal bone mineral density data of Bachrach \etal\ (1999)
 are not publicly available. Therefore we will illustrate
 fitting of additive mixed models using simulated data. 
 For simplicity we will use two ethnicity categories
 rather than four. \\
 
 \noindent
 Generate data:

\begin{small}
\begin{verbatim}
set.seed(394600) ; m <- 230 ; nVals <- sample(1:4,m,replace=TRUE)
betaVal <- 0.1 ; sigU <- 0.25 ; sigEps <- 0.05
f <- function(x)
    return(1 + pnorm((2*x-36)/5)/2)
U <- rnorm(m,0,sigU)
age <- NULL ; ethnicity <- NULL
Uvals <- NULL ; idNum <- NULL
for (i in 1:m)
{
   idNum <- c(idNum,rep(i,nVals[i]))
   stt <- runif(1,8,28-(nVals[i]-1))
   age <- c(age,seq(stt,by=1,length=nVals[i]))
   xCurr <- sample(c(0,1),1)
   ethnicity <- c(ethnicity,rep(xCurr,nVals[i]))
   Uvals <- c(Uvals,rep(U[i],nVals[i]))
}

epsVals <- rnorm(sum(nVals),0,sigEps)
SBMD <- f(age) + betaVal*ethnicity + Uvals + epsVals
\end{verbatim}
\end{small}  

\noindent 
Set up basic variables for the spline component:

\begin{small}
\begin{verbatim}
a <- 8 ; b <- 28; numIntKnots <- 15 
intKnots <- quantile(unique(age),seq(0,1,length=
                 (numIntKnots+2))[-c(1,(numIntKnots+2))])
\end{verbatim}
\end{small}    
   
\noindent
Obtain the spline component of the Z matrix.

\begin{small}
\begin{verbatim}
B <- bs(age,knots=intKnots,degree=3,
        Boundary.knots=c(a,b),intercept=TRUE)
Omega <- formOmega(a,b,intKnots)
eigOmega <- eigen(Omega)
indsZ <- 1:(numIntKnots+2)
UZ <- eigOmega$vectors[,indsZ]
LZ <- t(t(UZ)/sqrt(eigOmega$values[indsZ]))     
ZSpline <- B%*%LZ   
\end{verbatim}
\end{small}    

\noindent
Obtain the {\tt X} matrix:

\begin{small}
\begin{verbatim}
X <- cbind(rep(1,length(SBMD)),age,ethnicity)
\end{verbatim}
\end{small}   
   
\noindent
 Set up variables required for fitting via {\tt lme()}.
 Note that the random intercept is taken care
 of via the tree identification numbers variable `idNum',
 and that explicit formation of the random effect 
 contribution to the Z matrix is not required.

\begin{small}
\begin{verbatim}
groupVec <- factor(rep(1,length(SBMD)))
ZBlock <- list(list(groupVec=pdIdent(~ZSpline-1)),list(idNum=pdIdent(~1)))
ZBlock <- unlist(ZBlock,recursive=FALSE)
dataFr <- groupedData(SBMD~ethnicity|groupVec,
                      data=data.frame(SBMD,X,ZSpline,idNum))
fit <- lme(SBMD~-1+X,data=dataFr,random=ZBlock)
betaHat <- fit$coef$fixed
uHat <- unlist(fit$coef$random)
uSplineHat <- uHat[1:ncol(ZSpline)]
\end{verbatim}
\end{small}   

\noindent
Plot the data and fitted curve estimates together:

\begin{small}
\begin{verbatim}
ng <- 101 ; ageg <- seq(a,b,length=ng)
Bg <- bs(ageg,knots=intKnots,degree=3,Boundary.knots=c(a,b),intercept=TRUE)
ZgSpline <- Bg%*%LZ

plotMatrix0 <- cbind(rep(1,ng),ageg,rep(0,ng),ZgSpline)    
fhatgREML <- plotMatrix0 %*% c(betaHat, uSplineHat)

xLabs <- paste("ethnicity =",as.character(ethnicity))
pobj <- xyplot(SBMD~age|xLabs,groups=idNum,xlab="age (years)",
        ylab="spinal bone mineral density",subscripts=TRUE,
        panel=function(x,y,subscripts,groups)
        {
           panel.grid()  ; panel.superpose(x,y,subscripts,groups,
                                   type="b",col="grey60",pch=16)
           panelInd <- any(ethnicity[subscripts]==1)
           panel.xyplot(ageg,fhatgREML+panelInd*betaHat[3],
                        lwd=3,type="l",col="black")
        })
print(pobj)
\end{verbatim}
\end{small}   
   
\noindent
Print approximate 95\% confidence intervals for key parameters:
        
\begin{small}
\begin{verbatim}
print(intervals(fit))
\end{verbatim}
\end{small}

\begin{figure}[ht]
\null\centerline{\includegraphics[width=12cm]{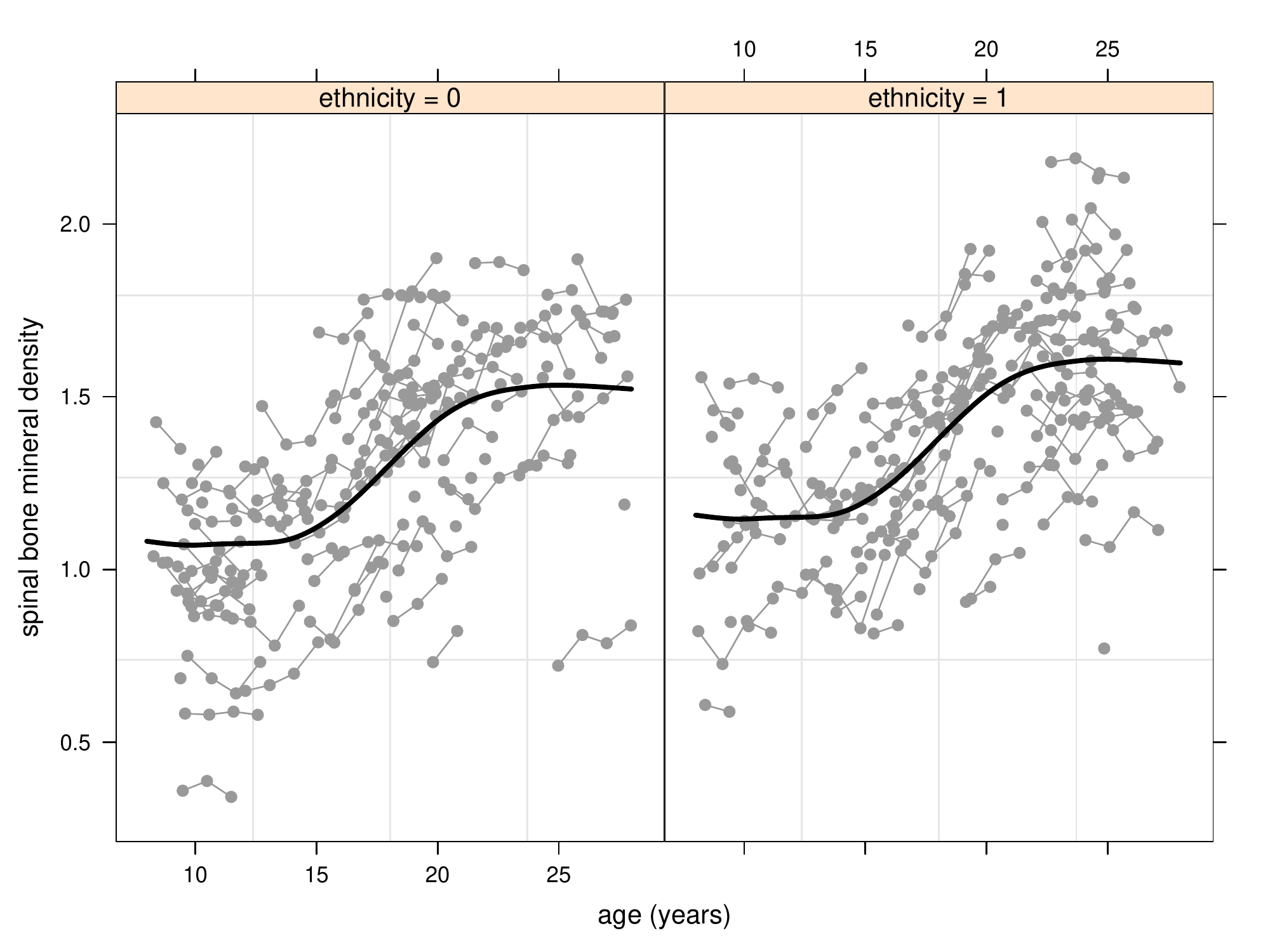}}
\caption
{\it
Plot obtained from execution of the last chunk
of code in this Appendix.
}
\label{fig:sitkaFit}
\end{figure}

\noindent
Execution of the above code should lead to 
an outcome similar to Figure \ref{fig:sitkaFit}
(the simulated data may differ between platforms).

\section*{References}

\BachrachHastieWangNarasimhanMarcus{1999}

\Berndt{1991}

\CarrollRuppertStefanskiCrainiceanu{2006}

\CantoniHastie{2002}

\ChaudhuriMarron{1999}

\CrainiceanuRuppertWand{2005}

\deBoor{1978}

\DenisonHolmesMallickSmith{2002}

\EilersMarx{1996}

\EubankTAS{1994}

\EubankBOOK{1999}

\GreenSilverman{1994}

\GuANOVA{2002}

\HallOpsomer{2005}

\HannigMarron{2006}

\HastieCRAN{2006}

\HastieTibshiraniFriedman{2001}

\LuoWahba{1997}

\NgoWand{2003}

\Nussbaum{1985}

\NychkaCummins{1996}

\OSullivanStatSci{1986}

\RuppertKNOTS{2002}

\RuppertWandCarroll{2003}

\SchoenbergGRAD{1964}

\Solo{2000}

\Speed{1991} 

\SpiegelhalterThomasBest{2000}

\Verbyla{1994}

\WahbaSIAM{1990}

\WandREGSPL{2000}

\WelhamCullisKenwardThompson{2007}

\WhittakerRobinson{1967}

\ZhaoStaudenmayerCoullWand{2006}

\end{document}